\documentclass[12pt]{iopart}
\usepackage{latexsym}
\usepackage{iopams}
\usepackage{setstack}
\usepackage{cite}
\newcommand{\be}{\begin{equation}}
\newcommand{\ee}{\end{equation}}
\begin{document}

\title{Electrostatic boundary value problems in the Schwarzschild background}

\author{P\'al G Moln\'ar}
\address{Institut f\"ur Theoretische Physik I, 
         Ruhr-Universit\"at Bochum, 
         D-44780 Bochum, Germany}


\begin{abstract}
The electrostatic potential of any test charge distribution in
Schwarzschild space with boundary values is derived. We calculate the Green's function,
generalize the second Green's identity for $p$-forms and find the
general solution. Boundary value problems are solved. With a multipole expansion the asymptotic property for the field of
any charge distribution is derived. It is shown that one produces a Reissner--Nordstr\"om black hole if one lowers a test
charge distribution slowly toward the horizon. The symmetry of the distribution is not important. All the multipole
moments fade away except the monopole. A calculation of the gravitationally induced electrostatic self-force on a
pointlike test charge distribution held stationary outside the black hole is presented.
\end{abstract}

\pacs{0240K, 0440N, 0470B, 4120C}

\maketitle

\section{Introduction}
In 1968, Israel proved the following theorem \cite{WI}: the Reissner--Nordstr\"om solution is the only static, asymptotically flat, electrovac
solution of Einstein's equations for which the surfaces $g_{00}=$ constant
are closed and simply connected and the event horizon $g_{00}=0$ is
regular. However, what happens, if a test charge is located near a Schwarzschild
black hole? The electrostatic field of a point charge at rest in Schwarzschild
space was derived by Hanni and Ruffini \cite{HR1} and Cohen and Wald \cite{CW} for multipole fields and by Linet \cite{BL} in algebraic form.
As the charge was slowly lowered into the hole, the authors found that the electric field of the charge remains well
behaved, while all the multipole moments, except the monopole, fade away. From Israel's theorem they concluded that a
Reissner--Nordstr\"om black hole was produced. Then in the early 1970s, when the black-hole uniqueness theorems were proved (particularly the theorems of Carter \cite{BC}, Hawking \cite{SH} and Robinson \cite{DR}), it became clear that an isolated black hole cannot have an electromagnetic field unless it is endowed with a net electric charge.

The closely related problem of a magnetic field outside a compact magnetic star with a surface current was considered
for dipole fields by Ginzburg and Ozernoi \cite{GO}, and for multipole fields by Anderson and Cohen \cite{AC}.
Petterson \cite{JP} presented a calculation of the quasistatic axisymmetric magnetic field in a Schwarzschild
background at radii both inside and outside the radius of the source. The authors found that the magnetic field
vanishes for an observer at infinity when the source approaches the horizon. This is in accordance with a theorem by
Price \cite{RP}, which states that during the process of gravitational collapse, all electromagnetic multipole moments
of the collapsing matter must disappear, except the electric monopole moment. However, the electric and magnetic
fields become very large when the sources are very near the horizon. This fact is probably only a manifestation of the
breakdown of the quasistatic treatment, and not a real effect. A time-dependent treatment by de la Cruz {\it et all}
\cite{Cetal} supports this view. They let a spherical shell of matter collapse at the speed of light upon a fixed
magnetic dipole. The numerical result is that the magnetic field outside the shell decays to zero during the collapse.
Wald \cite{RW} found the same sort of behavior when he calculated the electromagnetic field of an electrostatic or
magnetostatic multipole of fixed strength placed at the centre of a massive, non-rotating, spherical shell. As the shell approaches its own Schwarzschild radius, all electrostatic and magnetostatic multipoles except the monopole decay to zero and the electromagnetic field remains finite on the shell.

Electric and magnetic fields in the vicinity of massive compact objects are of particular interest for the study of
black hole magnetospheres which give hope to understand the mechanism behind the enormous power outputs of active
galactic nuclei and double radio sources. I shall not attempt to review the numerous works that have been written
addressing black hole magnetospheres (see, e.g., \cite{PG} and references therein).

In section~\ref{elpottcd} we calculate the electrostatic potential of any test charge distribution in a Schwarzschild
background. For that, first, we calculate the Green's function for a spherical shell bounded by $r=a$ and $b$ and then,
we make use of the generalized second Green's identity for $p$-forms (see the appendix). With this result, we are able
to solve boundary value problems. An example is demonstrated in section~\ref{tsex}. As in the usual electrodynamics, we make
in section~\ref{mulexp} a multipole expansion. It shows that for an observer far away from the sources all the multipole
moments, except the monopole, go to zero as the charge approaches the horizon. This was known only for a point charge.
The asymptotic properties, which are valid for any charge distribution, are presented in section~\ref{Asymp}. A force is
necessary in order to hold at rest a test charge distribution. In section~\ref{force} I recapitulate the works on the force
and I show how one can recover their results with the Green's funktion method. There is a repulsive self-force and
this force has for all static pointlike test distributions the same form. It depends only on the total charge $Q$. The derivation of the generalized second Green's identity for $p$-forms on (pseudo-) Riemannian manifolds can be found in the appendix.

\section{Electrostatic potential of a test charge distribution}
\label{elpottcd}
We use the standard Schwarzschild coordinates with the metric
\be \label{met}
\rmd s^{2}=\left( 1-\frac{2m}{r}\right) \rmd t^{2}-\left( 1-\frac{2m}{r}\right) ^{-1}\rmd r^{2}-r^{2}\left( \rmd\vartheta^{2}+\sin^{2}\!\vartheta\,\rmd\varphi^{2}\right) \;.
\ee
We write Maxwell's equations with differential forms \cite{NS}
\be \label{max}
\rmd F=0\;,\quad \delta F=4\pi J\;.
\ee
The first equation in (\ref{max}) implies, by Poincar\'e's lemma, the (local) existence of a potential $A=A_{\mu}\rmd x^{\mu}$ with $F=\rmd A$. Then, the inhomogeneous Maxwell equations become
\be \label{imax}
\delta \rmd A=4\pi J\;.
\ee
Since the Maxwell equations are invariant under a gauge transformation of the potential $A$, we may make use of this gauge freedom and require the Lorentz condition $\delta A=0$. Thus, Maxwell's equation (\ref{imax}) may be written in the form
\be \label{imaxl}
\Box A=4\pi J\;,
\ee
where $\Box := \delta\circ \rmd+\rmd\circ\delta$ is the Laplace--Beltrami operator. In the static case, when the
functions $A_{\mu}$ are independent on $t$, the electrostatic potential $A_{t}$ decouples from $A_{i}\,(i=r,\vartheta
,\varphi )$. Since we are interested in the field of a static test charge distribution, we set the spacelike
components of the current equal zero $j_{i}=0\,(i=r,\vartheta ,\varphi )$. Then, we may take $A_{i}=0$ and obtain as
the only non-trivial equation $(\mu =t)$
\be \label{zgla}
\fl\left( 1-\frac{2m}{r}\right) \frac{1}{r^{2}}\,\partial_{r}\!\left[ r^{2}A_{t,r}\right] +\frac{1}{r^{2}\sin\vartheta}\,\partial_{\vartheta}\!\left[ \,\sin\vartheta\,A_{t,\vartheta}\,\right] +\frac{1}{r^{2}\sin^{2}\!\vartheta}\,A_{t,\varphi\varphi}=-4\pi j_{t}\;.
\ee
The comma in equation (\ref{zgla}) denotes an ordinary derivative.

As in the usual electrostatics, we need a Green's function in order to derive the general solution for any test charge distribution. Therefore, we consider two new $1$--forms $G=G_{t}\,\rmd t$ and $\delta_{D} =\delta_{t}\,\rmd t$ with
\be \label{imaxg}
\Box G=4\pi\,\delta_{D}\;.
\ee
We call $G$ the `Green form' and $\delta_{D}$ the `Dirac form'. $\delta_{t}$ shall be the $\delta$--function with the normalization $\int\!\ast\delta_{D} =1$. Thus
\be \label{dnorm}
\delta_{t}=\frac{1-\frac{2m}{r}}{r^{2}\sin\vartheta}\,\delta (r-r')\,\delta (\vartheta -\vartheta ')\,\delta (\varphi -\varphi ')\;.
\ee
First, we solve the homogeneous equation of (\ref{imaxg}). If we make the separation ansatz
\be \label{sep}
G_{t}({\bi x},{\bi x}')=\sum_{l,m} R_{l}(r,r')Y_{lm}(\vartheta ,\varphi )\;,
\quad {\bi x}=(r,\vartheta,\varphi )\;,
\ee
we obtain the following equation for $R_{l}(r,r')$:
\be \label{rad}
\left( 1-\frac{2m}{r}\right) \frac{\rmd}{\rmd r}\!\left[ r^{2}\,\frac{\rmd R_{l}}{\rmd r}\right] -l(l+1)R_{l}(r,r')=0\;.
\ee
The solutions of equation (\ref{rad}) have been obtained independently bei Israel \cite{WI} and by Anderson and Cohen
\cite{AC}. With the transformation $u=\frac{r}{m}-1$ and the new function $U_{l}(u)=\sqrt{(1+u)/(1-u)}R_{l}(u)$ equation (\ref{rad}) becomes
\be \label{radu}
\left( 1-u^{2}\right) U_{l}''(u)-2u\,U_{l}'(u)+\left[ l(l+1)-\frac{1}{1-u^{2}}\right] U_{l}(u)=0\;.
\ee
The solutions of (\ref{radu}) are the associated Legendre functions of the first and second kind $P^{1}_{l}(u)$ and $Q^{1}_{l}(u)$ \cite{MO}. In the following, we take the way of Cohen and Wald \cite{CW}. They give as the two linearly independent solutions of equation (\ref{rad})
\numparts
\begin{eqnarray}
\fl g_{l}(r)  =  \cases{
                 1&for $l=0$ \quad ({\rm Definition})\,,\\
                 \frac{2^{l}\,l!\,(l-1)!\,m^{l}}{(2l)!}(r-2m)\frac{\rmd                           P_{l}}{\rmd r}
                 \left( \frac{r}{m}-1\right) &for $l \neq 0$\,,\\}\\
\fl f_{l}(r)  =  -\frac{(2l+1)!}{2^{l}\,(l+1)!\,l!\,m^{l+1}}(r-2m)\frac{\rmd Q_{l}}{\rmd r}
\left( \frac{r}{m}-1\right) \;, \label{radlosb}
\end{eqnarray}
\endnumparts
where $P_{l}$ and $Q_{l}$ are the two types of Legendre functions \cite{MO}. Then, they note three properties of $g_{l}(r)$ and $f_{l}(r)$, that will be important for the following analysis:
\begin{enumerate}
\item[(I)]For $l=0,\,g_{0}(r)=1$ (by definition) and $f_{0}(r)=1/r$.
\item[(II)]For all $l$, as $r\rightarrow\infty$, the leading term of $g_{l}(r)$ is $r^{l}$, while the leading term of $f_{l}(r)=1/r^{l+1}$.
\item[(III)]As $r\rightarrow 2m$, $f_{l}(r)\rightarrow$ finite constant, but $\rmd f_{l}/\rmd r$ blows up as $\ln (1-2m\,r^{-1})$ for $l\neq 0$. Since $g_{l}(r)=(r-2m)\times$(polynomial in $r$), so, as $r\rightarrow 2m$, $g_{l}(r)\rightarrow 0$ as $(r-2m)$ for $l\neq 0$.
\end{enumerate}
Now, we are able to write the homogeneous solution of (\ref{imaxg})
\be \label{homlosa}
G_{t}({\bi x},{\bi x}')=\cases{
             \sum_{l,m}\left[ A_{lm}\,g_{l}(r)+B_{lm}\,f_{l}(r)\right]                            Y_{lm}(\vartheta ,\varphi )&for $r>r'$\,, \\
             \sum_{l,m}\left[ C_{lm}\,g_{l}(r)+D_{lm}\,f_{l}(r)\right]                            Y_{lm}(\vartheta ,\varphi )&for $r<r'$\;.\\}
\ee
The constants $A_{lm}$, $B_{lm}$, $C_{lm}$, $D_{lm}$ can be determined by the boundary conditions. Suppose that the
boundary surfaces are concentric spheres at $r=a$ and $b$. The vanishing of $G_{t}({\bi x},{\bi x}')$ for ${\bi x}$ on the surface implies
\numparts
\begin{eqnarray}
C_{lm}\,g_{l}(a)+D_{lm}\,f_{l}(a)=0 \quad &{\rm for} \quad r=a\;,\\
A_{lm}\,g_{l}(b)+B_{lm}\,f_{l}(b)=0 \quad &{\rm for} \quad r=b\;.
\end{eqnarray}
\endnumparts
With $d_{lm} := -C_{lm}/f_{l}(a)=D_{lm}/g_{l}(a)$ and $b_{lm} := -A_{lm}/f_{l}(b)=B_{lm}/g_{l}(b)$ equation (\ref{homlosa}) becomes
\be \label{homlosb}
\fl G_{t}({\bi x},{\bi x}')=\cases{ \sum_{l,m}b_{lm}\,[g_{l}(b)f_{l}(r)-f_{l}(b)g_{l}(r)]\,Y_{lm}(\vartheta ,\varphi )&for $r>r'$\,, \\
\sum_{l,m}d_{lm}\,[g_{l}(a)f_{l}(r)-f_{l}(a)g_{l}(r)]\,Y_{lm}(\vartheta ,\varphi )&for $r<r'$\;.\\}
\ee
By continuity of $G_{t}$ at $r=r'$ we obtain
\be \label{con}
b_{lm}\,[g_{l}(b)f_{l}(r')-f_{l}(b)g_{l}(r')]=d_{lm}\,[g_{l}(a)f_{l}(r')-f_{l}(a)g_{l}(r')]\;.
\ee
Let
\[
a_{lm} := \frac{b_{lm}}{g_{l}(a)f_{l}(r')-f_{l}(a)g_{l}(r')}=\frac{d_{lm}}{g_{l}(b)f_{l}(r')-f_{l}(b)g_{l}(r')}\,,
\]
then we have
\be \label{homlosc}
G_{t}({\bi x},{\bi x}')= \sum_{l,m}a_{lm}R_{l}(r,r')Y_{lm}(\vartheta ,\varphi )\;,
\ee
where
\be \label{radf}
\fl R_{l}(r,r')=\cases{{[g_{l}(a)f_{l}(r')-f_{l}(a)g_{l}(r')][g_{l}(b)f_{l}(r)-f_{l}(b)g_{l}(r)]}&for $r>r'$\,,\\ 
{[g_{l}(b)f_{l}(r')-f_{l}(b)g_{l}(r')][g_{l}(a)f_{l}(r)-f_{l}(a)g_{l}(r)]}&for $r<r'$\;.\\}
\ee
We set equations (\ref{homlosc}), (\ref{radf}) and (\ref{dnorm}) into (\ref{imaxg}) and multiply both sides by $r^{2}$. Thus,
\begin{eqnarray}
\fl\lefteqn{\sum_{l,m}a_{lm}\left[ \left( 1-\frac{2m}{r}\right) \frac{\rmd}{\rmd r}\left( r^{2}\frac{\rmd R_{l}}{\rmd r}\right) -l(l+1)R_{l}(r,r')\right] Y_{lm}(\vartheta ,\varphi ) } \hspace{4cm} \nonumber \\
 & & {}=-\frac{4\pi}{\sin\vartheta }\left( 1-\frac{2m}{r}\right) \delta (r-r')\delta (\vartheta -\vartheta ')\delta (\varphi -\varphi ') \label{zglb}\;.
\end{eqnarray}
In order to evaluate $a_{lm}$, we multiply (\ref{zglb}) by $\sin\!\vartheta \,Y_{l'm'}^{\ast}(\vartheta ,\varphi )$
and integrate over $\vartheta$ and $\varphi$. Using the orthogonality condition of the spherical harmonics
$Y_{lm}(\vartheta ,\varphi )$, we obtain
\be \label{zglc}
a_{lm}\left[ \frac{\rmd}{\rmd r}\left( r^{2}\frac{\rmd R_{l}}{\rmd r}\right) -\frac{l(l+1)}{1-\frac{2m}{r}}\,R_{l}\right] =-4\pi\,\delta (r-r')\,Y_{lm}^{\ast}(\vartheta ',\varphi ')\;.
\ee
Integration over an infinitesimal interval at $r'$ gives
\begin{eqnarray}
\fl -4\pi\,Y^{\ast}_{lm}(\vartheta ',\varphi ') & = & a_{lm}\left[ {r'}^{2}\left. \frac{\rmd R_{l}(r>r')}{\rmd r}\right| _{r=r'}-{r'}^{2}\left. \frac{\rmd R_{l}(r<r')}{\rmd r}\right| _{r=r'} \right] \nonumber \\
 & = & a_{lm}{r'}^{2}[g_{l}(a)f_{l}(b)-g_{l}(b)f_{l}(a)]\left[ g_{l}(r')\frac{\rmd f_{l}}{\rmd r}(r')-f_{l}(r')\frac{\rmd g_{l}}{\rmd r}(r')\right] \nonumber\\
 & = & a_{lm}{r'}^{2}[g_{l}(a)f_{l}(b)-g_{l}(b)f_{l}(a)]\,W(g_{l},f_{l},r')\;, \label{zgld}
\end{eqnarray}
where $W(g_{l},f_{l},r')$ is the Wronskian of $g_{l}$ and $f_{l}$ at $r'$. The Wronskian
\be \label{wron}
W(r)=\left| \begin{array}{ccc}
            u_{1} & \ldots & u_{n} \\
            u'_{1} & \ldots & u'_{n} \\
            \vdots & & \vdots \\
            u^{(n-1)}_{1} & \ldots & u^{(n-1)}_{n}
            \end{array} \right|
\ee
of $n$ linearly independent solutions $u_{1}(r),\ldots,u_{n}(r)$ of a linear differential equation of the form  $u^{(n)}+a_{n-1}(r)u^{(n-1)}+\ldots+a_{0}(r)u=b(r)$ satisfies \cite{WW}
\be \label{wronbed}
W(r)=W(r_{0})\exp\left( -\int^{r}_{r_{0}}a_{n-1}(s)\,\rmd s\right) \;.
\ee
In our case $a_{n-1}=a_{1}(r)=2/r$ and therefore
\be \label{wronlos}
W(g_{l},f_{l},r)=W(g_{l},f_{l},r_{0})\frac{r_{0}^{\,2}}{r^{\,2}}\;.
\ee
So, we have
\be \label{wroncon}
r^{\,2}W(g_{l},f_{l},r)={\rm constant}\;.
\ee
We find the constant by evaluating $W(g_{l},f_{l},r)$ for large values of $r$. Property (II) gives the values for the functions $g_{l}$ and $f_{l}$ for large $r$. Then, by equation (\ref{wroncon}), we have for all $r$
\begin{eqnarray}
r^{\,2}W(g_{l},f_{l},r) & = & r^{\,2}\left[ -r^{\,l}(l+1)\frac{1}{r^{\,l+2}}-\frac{1}{r^{\,l+1}}\,l\,r^{\,l-1}\right] \nonumber \\
 & = & -(2l+1)\;. \label{wronlc}
\end{eqnarray}
Setting (\ref{wronlc}) in (\ref{zgld}) and solving for $a_{lm}$ gives
\be \label{conlm}
a_{lm}=\frac{4\pi}{2l+1}\,\frac{1}{g_{l}(a)f_{l}(b)-g_{l}(b)f_{l}(a)}\,Y^{\ast}_{lm}(\vartheta ',\varphi ')\;.
\ee
Thus, the solution of (\ref{imaxg}) is
\begin{eqnarray}
G_{t}({\bi x},{\bi x}')=&4\pi \sum_{l,m}\frac{Y^{\ast}_{lm}(\vartheta ',\varphi
')Y_{lm}(\vartheta,\varphi)}{(2l+1)\left[ 1-(g_{l}(a)f_{l}(b))/(f_{l}(a)g_{l}(b))\right] } \nonumber\\
&\times\left( g_{l}(r_{<})-\frac{g_{l}(a)}{f_{l}(a)}f_{l}(r_{<})\right) \left( f_{l}(r_{>})-\frac{f_{l}(b)}{g_{l}(b)}g_{l}(r_{>})\right) \;, \label{grlos}
\end{eqnarray}
where $r_{<}$ ($r_{>}$) is the smaller (larger) of $r$ and $r'$. Now, we use the second Green's identity (equation (\ref{sGi}) in the appendix) and set $u := A(x)$, $v:=G(x,x')$ with $A=A_{t}\,\rmd t$ and $G=G_{t}\,\rmd t$. Since $G_{t}({\bi x},{\bi x}')$ vanishes on the surface $\partial D$, we obtain as the solution of (\ref{zgla})
\begin{eqnarray}
A_{t}({\bi x}) & = & \int_{D}j_{t}({\bi x}')\,G_{t}({\bi x},{\bi x}')\,\frac{r'^{\,2}\sin\vartheta '}{1-\frac{2m}{r'}}\,\rmd r'\rmd\vartheta '\rmd\varphi ' \nonumber\\
 & & {}-\frac{1}{4\pi}\int_{\partial D}A_{t}({\bi x}')\,\frac{\partial G_{t}({\bi x},{\bi x}')}{\partial r'}\,{r'}^{2}\sin\vartheta '\rmd\vartheta '\rmd\varphi '\;.  \label{potlos}
\end{eqnarray}
Since the functions are not dependent on time, we neglect the integration over $t$. One have to choose $j_{t}({\bi x})$ in (\ref{potlos}) in such a way, that
\be \label{normj}
\int_{D}j_{t}({\bi x})\,\frac{r^{\,2}\sin\vartheta}{1-\frac{2m}{r}}\,\rmd r\,\rmd\vartheta\, \rmd\varphi =Q\;,
\ee
where $Q$ is the total charge.

In 1976, Linet presented an algebraic solution for a point charge situated at the point $(r_0,\vartheta_0,\varphi_0)$ with
$r_0>2m$ \cite{BL}. He obtained his solution by modifying a particular analytic solution found by Copson \cite{EC}, which did not
satisfy the boundary condition. Linet's solution is
\begin{eqnarray}
\fl A_t^L({\bi x})=-\frac{em}{r_0r} \nonumber\\
\fl -\frac{e}{r_0r}\frac{(r-m)(r_0-m)-m^2\lambda (\vartheta ,\varphi )}{\left[
(r-m)^2+(r_0-m)^2-m^2-2(r-m)(r_0-m)\lambda (\vartheta ,\varphi )+m^2\lambda^2(\vartheta ,\varphi )\right] ^{1/2}} \label{atL}
\end{eqnarray}
with
\be \label{lam1}
\lambda (\vartheta ,\varphi )=\cos\vartheta\cos\vartheta_0+\sin\vartheta\sin\vartheta_0\cos (\varphi -\varphi_0)\;.
\ee
Now, it is easy to generalize equation (\ref{atL}) for any charge distribution $j_{t}({\bi x})$. With (\ref{sGi}) again
\be \label{atLG}
A_{t}^L({\bi x})=\int_{D}j_{t}({\bi x}')\,G_{t}^L({\bi x},{\bi x}')\,\frac{r'^{\,2}\sin\vartheta '}{1-\frac{2m}{r'}}\,\rmd
r'\rmd\vartheta '\rmd\varphi '\;,
\ee
\begin{eqnarray}
\fl G_t^L({\bi x},{\bi x}')=-\frac{m}{r'r} \nonumber\\
\fl -\frac{1}{r'r}\frac{(r-m)(r'-m)-m^2\lambda (\vartheta ,\varphi )}{\left[
(r-m)^2+(r'-m)^2-m^2-2(r-m)(r'-m)\lambda (\vartheta ,\varphi )+m^2\lambda^2(\vartheta ,\varphi )\right] ^{1/2}}\;,\label{GtL}
\end{eqnarray}
\be \label{lam2}
\lambda (\vartheta ,\varphi )=\cos\vartheta\cos\vartheta'+\sin\vartheta\sin\vartheta'\cos (\varphi -\varphi')\;.
\ee
The boundary values are equal zero. In order to consider some boundary values, one should add to (\ref{GtL}) a solution of the
homogeneous equation, which is fully determined by the choice of the boundary conditions.

\section{Solution of a boundary value problem}
\label{tsex}
As an example let us consider a concentric ring of charge of radius $R>2m$ and total charge $Q$ located at $\vartheta
'=\frac{1}{2}\,\pi$ inside a hollow sphere of radius $b>R$ with the potential $V(\vartheta ',\varphi ')$ on the surface. With $a\rightarrow 2m$ in (\ref{grlos}), the derivative of $G_{t}$, evaluated at $r'=b$, is
\be \label{pktvert}
\fl\left. \frac{\partial G_{t}}{\partial r'}\right|_{r'=b}=4\pi\sum_{l,m}\frac{Y^{\ast}_{lm}(\vartheta ',\varphi ')Y_{lm}(\vartheta,\varphi)}{(2l+1)}\,g_{l}(r)\left( \left. \frac{\rmd f_{l}(r')}{\rmd r'}\right|_{r'=b}-\frac{f_{l}(b)}{g_{l}(b)}\left. \frac{\rmd g_{l}(r')}{\rmd r'}\right|_{r'=b}\right) \;.
\ee
Consequently, the potential of the hollow sphere inside $r=b$ is, according to (\ref{potlos}),
\begin{eqnarray}
\Phi ({\bi x})=&-\sum_{l,m}\left[ \int V(\vartheta ',\varphi ')Y^{\ast}_{lm}(\vartheta ',\varphi ')\sin \vartheta '\rmd\vartheta '\rmd\varphi '\right] \nonumber\\
&\times\frac{b^{2}g_{l}(r)}{2l+1}\left( \left. \frac{\rmd f_{l}(r')}{\rmd r'}\right|_{r'=b}-\frac{f_{l}(b)}{g_{l}(b)}\left. \frac{\rmd g_{l}(r')}{\rmd r'}\right|_{r'=b}\right) Y_{lm}(\vartheta ,\varphi )\;. \label{pktlos}
\end{eqnarray}
The charge density of the ring can be read off from (\ref{normj}) 
\be \label{ringvert}
j_{t}=\frac{Q}{2\pi}\,\frac{1-\frac{2m}{r}}{r^{\,2}\sin\vartheta}\,\delta (r-R)\,\delta (\vartheta
-{\textstyle\frac{1}{2}}\,\pi)\;.
\ee
Since the problem is axially symmetric, we set $m=0$. Then, using (\ref{grlos}) with $a\rightarrow 2m$, equation (\ref{potlos}) becomes
\be \label{ringlos}
\fl A_{t}({\bi x})=Q\sum_{n=0}^{\infty}(-1)^{n}\,\frac{(2n)!}{2^{2n}(n!)^{2}}\,g_{2n}(r_{<})\left[ f_{2n}(r_{>})-\frac{f_{2n}(b)}{g_{2n}(b)}\,g_{2n}(r_{>})\right] P_{2n}(\cos\vartheta )+\Phi({\bi x})\;,
\ee
where we have used \cite{MO}
\be \label{moform}
P_{l}(0)=\cases{(-1)^{n}\,\frac{(2n)!}{2^{2n}(n!)^{2}}&for $l=2n$\,,\\
                0&for $l=2n+1$\;.\\}
\ee
$r_{<}$ ($r_{>}$) is the smaller (larger) of $r$ and $R$.

\section{Multipole expansion}
\label{mulexp}
A located distribution of charge $j_{t}$ is non-vanishing only inside a sphere of radius $R>2m$ around the origin and
the charge is in the space between $r=2m$ and $R$. Since we are interested in the moments outside the charge
distribution, $r_{<}=r'$ and $r_{>}=r$. With $a\rightarrow 2m$, $b\rightarrow\infty$ in (\ref{grlos}) and the
vanishing of the potential on the boundary surface in (\ref{potlos}) we find the following multipole expansion for the
potential $A_{t}$:
\be \label{potexp}
A_{t}=4\pi \sum_{l,m}\frac{1}{2l+1}\,f_{l}(r)\,q^{\ast}_{lm}\,Y_{lm}(\vartheta ,\varphi )\;,
\ee
where the multipole moments $q_{lm}$ are defined by
\be \label{mumo}
q_{lm}=\int g_{l}(r')\,Y_{lm}(\vartheta ',\varphi ')\,j_{t}(r',\vartheta ',\varphi ')\,\frac{r'^{\,2}\sin\vartheta
'}{1-2m/r'}\,\rmd r'\,\rmd\vartheta '\,\rmd\varphi '\;.
\ee
We give only the moments with $m\geqslant 0$, since the moments with $m<0$ are related through
\be \label{mumorel}
q_{l-m}=(-1)^{m}q^{\ast}_{lm}\;.
\ee
For $l=0$ and $l=1$ equation (\ref{mumo}) becomes
\be \label{mumo0}
q_{00}=\frac{Q}{\sqrt{4\pi}}\;,
\ee
\be \label{mumo1}
\eqalign{q_{11}&=-\sqrt{\frac{3}{8\pi}}\,(p_{x}+ip_{y})\;,\\
q_{10}&=\sqrt{\frac{3}{4\pi}}\,p_{z}\;,}
\ee
where
\be \label{dimo}
{\bi p}=\int \frac{g_{1}(r)}{r}\,{\bi x}\,j_{t}({\bi x})\,\frac{r^{\,2}\sin\vartheta }{1-2m/r}\,\rmd r\,\rmd\vartheta \,\rmd\varphi
\ee
is the electric dipole moment and ${\bi x}=(x,y,z)$ with $x=r\sin\vartheta \cos\varphi$, $y=r\sin\vartheta \sin\varphi $, $z=r\cos\vartheta $. For $l=2$ we have
\be \label{mumo2}
\eqalign{
q_{22}&=\frac{1}{12}\,\sqrt{\frac{15}{2\pi}}\,(Q_{11}+2i\,Q_{12}-Q_{22})\;,\\
q_{21}&=-\frac{1}{3}\,\sqrt{\frac{15}{8\pi}}\,(Q_{13}+i\,Q_{23})\;,\\
q_{20}&=\frac{1}{2}\,\sqrt{\frac{5}{4\pi}}\,Q_{33}\;,}
\ee
where $Q_{ij}$ is the traceless quadrupole moment tensor
\be \label{qamo}
Q_{ij}=\int \frac{g_{2}(r)}{r^{\,2}}\,(3x_{i}x_{j}-\delta_{ij}r^{\,2})\,j_{t}({\bi x})\,
\frac{r^{\,2}\sin\vartheta }{1-2m/r}\,\rmd r\,\rmd\vartheta \,\rmd\varphi \;.
\ee
With equations (\ref{mumo0}), (\ref{mumo1}) and (\ref{mumo2}) the expansion of $A_{t}({\bi x})$ (\ref{potexp}) begins
\be \label{muexp}
A_{t}({\bi x})=\frac{Q}{r}+\frac{f_{1}(r)}{r}\,{\bi p}\cdot{\bi
x}+\frac{1}{2}\,\frac{f_{2}(r)}{r^{\,2}}\,\sum_{i,j}Q_{ij}x_{i}x_{j}+\cdots\,.
\ee

Now, we go back to equation (\ref{potexp}). The electric field components for a given multipole can be expressed most easily in terms of spherical coordinates. In the local orthonormal basis
\be \label{onba}
\fl\theta^{0}=\sqrt{1-\frac{2m}{r}}\,\rmd t\,,\quad \theta^{1}=\frac{1}{\sqrt{1-2m/r}}\,\rmd r\,,\quad \theta^{2}=r\,\rmd\vartheta\,,\quad \theta^{3}=r\,\sin\!\vartheta\,\rmd\varphi\,,
\ee
the components of the field tensor $F=\rmd A$ with $A=A_{t}\,\rmd t$ are
\be \label{feldten}
\fl
F=-A_{t,r}\,\theta^{0}\wedge\theta^{1}-A_{t,\vartheta}\,\frac{1}{\sqrt{1-2m/r}}\,\frac{1}{r}\,\theta^{0}\wedge\theta^{2}-A_{t,\varphi}\,\frac{1}{\sqrt{1-2m/r}}\,\frac{1}{r\sin\vartheta}\,\theta^{0}\wedge\theta^{3}\;,
\ee
\be \label{feldkomp}
\fl F_{01}\equiv E_{r}\,,\quad F_{02}\equiv E_{\vartheta}\,,\quad F_{03}\equiv E_{\varphi}\;.
\ee
The electric field with definite $l,m$ has spherical components
\be \label{feldlm}
\eqalign{
E_{r}&=-\frac{4\pi}{2l+1}\,q^{\ast}_{lm}\,\frac{\rmd f_{l}(r)}{\rmd r}\,Y_{lm}(\vartheta ,\varphi )\;,\\
E_{\vartheta}&=-\frac{4\pi}{2l+1}\,q^{\ast}_{lm}\,\frac{f_{l}(r)}{r\sqrt{1-2m/r}}\,\frac{\partial}{\partial\vartheta}Y_{lm}(\vartheta ,\varphi )\;,\\
E_{\varphi}&=-\frac{4\pi}{2l+1}\,q^{\ast}_{lm}\,\frac{f_{l}(r)}{r\sqrt{1-2m/r}}\,\frac{im}{\sin\vartheta}\,Y_{lm}(\vartheta ,\varphi )\;.}
\ee
For a dipole ${\bi p}$ along the $z$ axis, the fields in (\ref{feldlm}) reduce to the form
\be \label{felddi}
\fl E_{r}=-p\,\frac{\rmd f_{1}(r)}{\rmd r}\,\cos\vartheta\,,\quad E_{\vartheta}=p\,\frac{f_{1}(r)}{r\sqrt{1-2m/r}}\,\sin\vartheta\,,\quad 
E_{\varphi}=0\;.
\ee

\section{Asymptotic fields for $\lowercase{r}'\rightarrow 2\lowercase{m}$}
\label{Asymp}
Cohen and Wald \cite{CW} showed that, if one lowers a point test charge slowly toward $r'=2m$, one produces a
Reissner--Nordstr\"om black hole. Is this true for any charge distribution? First, we consider the case of a charge
distribution at the point $r'>2m$ and an observer at $2m\leqslant r<r'$. Thus, $r_{<}=r$ and $r_{>}=r'$. Setting equations (\ref{grlos}) and (\ref{potlos}) (with $a\rightarrow 2m$, $b\rightarrow\infty$) in (\ref{feldten}) the components of the field in the orthonormal basis (\ref{onba}) become
\be \label{feldexp}
\eqalign{
F_{01}&=-4\pi\sum_{l,m}\frac{1}{2l+1}\,\frac{\rmd g_{l}(r)}{\rmd r}\,Y_{lm}(\vartheta ,\varphi )\,\mu_{lm}\;,\\
F_{02}&=-\frac{4\pi}{r\sqrt{1-2m/r}}\,\sum_{l,m}\frac{1}{2l+1}\,g_{l}(r)\,\frac{\partial Y_{lm}(\vartheta ,\varphi )}{\partial\vartheta }\,\mu_{lm}\;,\\
F_{03}&=-\frac{4\pi}{r\sin\vartheta\sqrt{1-2m/r}}\,\sum_{l,m}\frac{1}{2l+1}\,g_{l}(r)\,\frac{\partial Y_{lm}(\vartheta ,\varphi )}{\partial\varphi }\,\mu_{lm}\;,}
\ee
where
\be \label{moexp}
\mu_{lm}=\int j_{t}(r',\vartheta ',\varphi ')\,f_{l}(r')\,Y^{\ast}_{lm}(\vartheta ',\varphi
')\,\frac{r'^{\,2}\sin\vartheta '}{1-2m/r'}\,\rmd r'\,\rmd\vartheta '\,\rmd\varphi '\;.
\ee
For $r$ near $2m$, we see, by property (III) of the functions $g_{l}$, that $F_{01}$ remains finite and $F_{02}\sim
F_{03}\sim \Or [(1-2m/r)^{1/2}]$. A stationary observer at $r'>r\approx 2m$ sees a radial electrostatic field. Since
$j_{t}$ can always have a term $(1-2m/r)/(r^{\,2}\sin\vartheta )$ (cf~(\ref{normj})), the corresponding term in (\ref{moexp}) vanishes. So, we see, by property (III) of the functions $f_{l}$, that the field components $F_{0i}$ ($i=1,2,3$) remain finite at $r=2m$ as $r'\rightarrow 2m$.

In the case $r>r'$, we make use of equations (\ref{feldlm}) and (\ref{mumo}). As seen above, the term
$(r'^{\,2}\sin\vartheta ')/(1-2m/r')$ in (\ref{mumo}) vanishes with the help of the corresponding term in $j_{t}$. From the fact that, for all $l\neq 0$, $g_{l}(r')\rightarrow 0$ as $r'\rightarrow 2m$ (property (III)), we conclude that all the multipole moments of $q_{lm}$ (\ref{mumo}) except the monopole go to zero as $r'\rightarrow 2m$. Since $q_{00}=Q/\sqrt{4\pi}$ and $f_{0}(r)=r^{-1}$, we have the result that for all $r>2m$
\be \label{rnv}
E_{r}=\frac{Q}{r^{\,2}}\,,\quad E_{\vartheta}=E_{\varphi}=0\,,\quad {\rm as}\quad r'\rightarrow 2m\;.
\ee
Thus, although the charge distribution does not possess any symmetry, the electrostatic field approaches the spherically symmetric Reissner--Nordstr\"om value $E_{r}=Q/r^{\,2}$ for $r'\rightarrow 2m$.

\section{Force on a charge distribution}
\label{force}
One may ask what force is necessary to hold at rest a test charge distribution at a point $(r_0,\vartheta_0,\varphi_0)$
outside a Schwarzschild black hole in a freely falling local system. A number of authors \cite{DeW,BG,CMac,AV} have investigated
the problem by assuming that the gravitational field was weak; thus they worked to leading order in the small quantity
$m/r$. Smith and Will \cite{SW} presented an exact calculation for a charged test particle held stationary near a
Schwarzschild black hole. Zel'nikov and Frolov considered the influence of the gravitational field of a charged black hole
on the self-energy of a charged particle \cite{ZF1} and found that there is a mass shift of the particle in the
gravitational field and that the absolute value of the mass shift coincides with the absolute value of the shift for a
uniformly accelerated electron \cite{ZF2}.

Smith and Will \cite{SW} got the result that the force is given by two terms. The first term is just the negative of the
gravitational field that the hole exerts on the test particle. The second term is the gravitationally induced self-force
of the particle. The self-force is repulsive and has the magnitude (in Schwarzschild coordinates)
\be \label{fself}
F_{{\rm self}}=\frac{me^2}{r^3}\;.
\ee
Since the hole is uncharged and the self--force vanishes as $m\rightarrow 0$, we must assume that the effect is induced by
the spacetime curvature. The gravitational field obviously modifies the elctrostatic self-interaction of the charged
particle in such a way that the particle experiences a finite self-force.

For our calculation with the test charge distribution we use what was called the `global method' by Smith and Will in
their paper \cite{SW}. If we displace the charge slowly by a distance $\delta{\bi x}_0$ toward the hole then, according to the
freely falling system, an amount of work $\delta\overline{W}$ is done given in this system by
\be \label{arb}
\delta\overline{W}=-\overline{\bi F}_{\rm ext}\cdot\delta\overline{{\bi x}}_0\;.
\ee
Because of the gravitational red-shift, the energy $\delta E$ received by an observer at asymptotic infinity is
\be \label{ene}
\delta E=\sqrt{g_{00}(r_0)}\,\delta\overline{W}\;.
\ee
However, conservation of energy forces this energy to coincide with the change in the asymptotically measured mass
$-\delta M$ of the system. In the freely falling system the coordinates can be chosen to be locally flat; that means
\be \label{locfl}
g_{\,\overline{\alpha\beta}}=\eta_{\,\overline{\alpha\beta}}\,,\quad
\partial_{\overline{\gamma}}\,g_{\,\overline{\alpha\beta}}=0\;.
\ee
The transformation between the arbitrary coordinates and the locally flat ones is done by
\be \label{transf}
g_{\mu\nu}=\eta_{\,\overline{\alpha\beta}}\,\Lambda^{\overline{\alpha}}_{\,\,\,\,\mu}\,\Lambda^{\overline{\beta}}_{\,\,\,\,\nu}\;.
\ee
Hence, we have
\be \label{trco}
\delta x^{\overline{\alpha}}=\Lambda^{\overline{\alpha}}_{\,\,\,\,\beta}\,\delta x^\beta\;.
\ee
Then, the external force on the test distribution in the locally flat system is
\be \label{exfo}
F^{\overline{\mu}}_{\rm ext}=\frac{1}{\sqrt{g_{00}(r_0)}}\,\frac{\delta M}{\Lambda^{\overline{\mu}}_{\,\,\,\,\nu}\,\delta
x^\nu_0}\;.
\ee
The change in mass $\delta M$ between two nearby, non-rotating black--hole configurations was calculated by Carter
\cite{BC2}
\be \label{blho}
\delta M=\frac{\varkappa}{8\pi}\,\delta A-\frac{1}{8\pi}\,\delta\!\int G^0_{\,\,0}\sqrt{-g}\,\rmd^3x+\frac{1}{16\pi}\,\int
G^{\mu\nu}h_{\mu\nu}\sqrt{-g}\,\rmd^3x\;,
\ee
where $\varkappa$ and $A$ are the surface gravity and the area of the black hole, respectively, $G^{\mu\nu}$ is the
Einstein tensor, and $h_{\mu\nu}$ is the difference in the metric between the two configurations. The integrals are to be
evaluated over the exterior of the black hole. In our case, we can set $\delta A=0$ and ignore the term involving
$h_{\mu\nu}$. Thus, with Einstein's equations, we have
\be \label{chma}
\delta M=-\delta\!\int T^0_{\,\,\,\,0}\sqrt{-g}\,\rmd^3x\;.
\ee
Since the energy--momentum tensor $T^{\mu\nu}$ of the system has a mechanical contribution and an electromagnetic
contribution, we split the `energy' integral above into two terms
\be \label{eneint}
-\int T^0_{\,\,\,\,0}\sqrt{-g}\,\rmd^3x\equiv U_{\rm mech}+U_{\rm em}\;.
\ee
With \eref{chma} and \eref{eneint} equation \eref{exfo} becomes
\be \label{exfo2}
F^{\overline{\mu}}_{\rm ext}=\frac{1}{\sqrt{g_{00}(r_0)}}\,\frac{\delta (U_{\rm mech}+U_{\rm em})}{\Lambda^{\overline{\mu}}_{\,\,\,\,\nu}\,\delta
x^\nu_0}\;.
\ee
To evaluate the mechanical contribution, we must choose a particle model. The easiest one is the model of an ideal fluid
\be \label{idfl}
T^{\mu\nu}=(\varrho^0+p)\,u^\mu u^\nu -p\,g^{\mu\nu}\;,
\ee
where $p$ is the pressure, $\varrho^0$ the density and $u^\mu$ the velocity field with
\be \label{norm}
g_{\mu\nu}u^\mu u^\nu =1\;.
\ee
Since for the particle density $\varrho^0$
\be \label{toma}
\int \varrho^0 \sqrt{-g}\,\rmd^3x=m_0\;,
\ee
where $m_0$ is the total mass of the charge density, we obtain
\be \label{teidi}
\varrho^0=\frac{m_0}{\sqrt{-g}}\,\delta^3({\bi x}-{\bi x}_0)\;.
\ee
In our system $p=0$, thus
\be \label{idfl2}
T^{\mu\nu}=\frac{m_0}{\sqrt{-g}}\,\frac{\rmd x^\mu}{\rmd\tau}\,\frac{\rmd x^\nu}{\rmd\tau}\,\delta^3({\bi x}-{\bi x}_0)\;,
\ee
where $x^\mu (\tau)$ is the worldline of the charge density and $\tau$ the proper time. We write \eref{idfl2} in
manifestly covariant form
\be \label{idfl3}
T^{\mu\nu}=\frac{m_0}{\sqrt{-g}}\int\frac{\rmd x^\mu}{\rmd\tau}\,\frac{\rmd
x^\nu}{\rmd\tau}\,\delta^4(x-x_0(\tau))\rmd\tau\;.
\ee
In the local flat system $u^{\overline{\mu}}=(1,0,0,0)$. One finds from \eref{trco} that
\be \label{prti}
\rmd\tau =\sqrt{g_{00}}\,\rmd t\;.
\ee
Choosing $\tau =0$ when $t=0$, we have
\be \label{weli}
x^{\mu}(\tau)=\left( \frac{1}{\sqrt{g_{00}}}\,\tau,x_0,y_0,z_0\right) \;.
\ee
Now, the mechanical contribution to the energy is easily found from \eref{idfl3} and \eref{weli} to be
\be \label{meco}
U_{\rm mech}=-m_0\sqrt{g_{00}(r_0)}\;.
\ee

To evaluate the electromagnetic contribution, we first note that
\be \label{tem}
T^{\mu\nu}_{\rm em}=\frac{1}{4\pi}\,\left[
F^\mu_{\,\,\,\,\lambda}F^{\lambda\nu}+\frac{1}{4}\,g^{\mu\nu}F^{\sigma\lambda}F_{\sigma\lambda}\right] \;.
\ee
So, it follows that
\be \label{tem2}
T_{\rm em}{}^{0}_{\,\,0}=-\frac{1}{8\pi}\,g^{00}g^{ij}\partial_iA_0\,\partial_jA_0
\ee
and hence
\be \label{emen}
U_{\rm em}=\frac{1}{8\pi}\,\int g^{00}g^{ij}\partial_iA_0\,\partial_jA_0\sqrt{-g}\,\rmd^3x\;.
\ee
Via an integration by parts, one obtains
\begin{eqnarray}
\fl \lefteqn{U_{\rm
em}=-\frac{1}{8\pi}\,\int\rmd^3xA_0\,\partial_i(\sqrt{-g}\,g^{00}g^{ij}\partial_jA_0)+\frac{1}{8\pi}\,\int_{r\rightarrow\infty}\rmd^2S_i\,\sqrt{-g}\,g^{00}g^{ij}A_0\,\partial_jA_0}\nonumber\\
 & & -\frac{1}{8\pi}\,\int_{r=2m}\rmd^2S_i\,\sqrt{-g}\,g^{00}g^{ij}A_0\,\partial_jA_0\;. \label{emen2}
\end{eqnarray}
The surface integrals vanish because $A_0$ drops off as $1/r$ at large distances and the angular integration on the horizon
averages to zero as one can see using \eref{atL}. One is left with the first term, which because of \eref{max} becomes
\be \label{emen3}
U_{\rm em}=-\frac{1}{2}\int j^0A_0\sqrt{-g}\,\rmd^3x\;.
\ee
Now, we replace $A_0$ with the help of our Green's funktion $G^L_t({\bi x},{\bi x}')$ (equation \eref{GtL})
\be \label{emen4}
U_{\rm em}=-\frac{1}{2}\int j^0({\bi x})\sqrt{-g({\bi x})}\,\rmd^3x\int j^0({\bi x}')\sqrt{-g({\bi x}')}\,\rmd^3x'\,G^L_t({\bi x},{\bi x}')\;.
\ee
In the integral above, $G^L_t({\bi x},{\bi x}')$ corresponds to the energy integral of a point test charge at ${\bi
x}={\bi x}'$. We can this clarify by setting $j^0({\bi x}')=\int\rmd^3x''\sqrt{-g({\bi x}'')}\,\delta^0({\bi x}'-{\bi
x}'')j^0({\bi x}'')$. Then, equation \eref{emen4} becomes
\be \label{emen5}
U_{\rm em}=\int j^0({\bi x})\sqrt{-g({\bi x})}\,\rmd^3x\int j^0({\bi x}'')\sqrt{-g({\bi x}'')}\,\rmd^3x''\,U_{\rm
em}^\delta
\ee
with
\be \label{empken}
U_{\rm em}^\delta=-\frac{1}{2}\int\rmd^3x'\sqrt{-g({\bi x}')}\,\delta^0({\bi x}'-{\bi x}'')\,G^L_t({\bi x},{\bi x}')\;.
\ee
$G^L_t({\bi x},{\bi x}')$ has a singularity at the point ${\bi x}={\bi x}'$. In order to separate out the divergent
behavior, we expand $G^L_t({\bi x},{\bi x}')$ about ${\bi x}={\bi x}'$. We set $|{\bi x}-{\bi x}'|=a$ and take the limit
$a\rightarrow 0$ after the integration. This corresponds to giving the point a finite radius $a$. For the integration it
proves convenient to write $G^L_t({\bi x},{\bi x}')$ in isotropic coordinates
\be \label{isco}
\eqalign{\varrho =\textstyle\frac{1}{2}\left[ r-m+(r^2-2mr)^{1/2}\right] \;,\\
x=\varrho\sin\vartheta\cos\varphi\;,\\
y=\varrho\sin\vartheta\sin\varphi\;,\\
z=\varrho\cos\vartheta\;.}
\ee
Then
\be \label{scis}
r=\varrho\left( 1+\frac{m}{2\varrho}\right) ^{1/2}
\ee
and the Schwarzschild metric has the form
\be \label{isme}
g=h^2(|{\bi x}|)\,\rmd t^2-f^2(|{\bi x}|)\,\rmd{\bi x}^2
\ee
with
\be \label{iscoe}
h(\varrho )=\frac{1-m/2\varrho}{1+m/2\varrho}\,,\quad f(\varrho )=\left( 1+\frac{m}{2\varrho}\right) ^2\;.
\ee
After a little calculation $G^L_t({\bi x},{\bi x}')$ (equation \eref{GtL}) becomes
\begin{eqnarray}
\fl G^L_t({\bi x},{\bi x}')=-\frac{1}{\varrho '(1+m/2\varrho ')^2\varrho (1+m/2\varrho )^2} \nonumber\\
\lo\times\left[ m+\varrho '\left( \frac{\varrho^2-\frac{m^2}{2\varrho
'^2}(x'x+y'y+z'z)+\tilde{\varrho}'^2}{\varrho^2-2(x'x+y'y+z'z)+\varrho '^2}\right) ^{1/2}\right. \nonumber\\
\left. \lo+\tilde{\varrho}'\left(
\frac{\varrho^2-2(x'x+y'y+z'z)+\varrho '^2}{\varrho^2-\frac{m^2}{2\varrho
'^2}(x'x+y'y+z'z)+\tilde{\varrho}'^2}\right) ^{1/2}\right] \;, \label{isGtL}
\end{eqnarray}
where $\tilde{\varrho}=m^2/4\varrho$. For ${\bi x}'=(0,0,b)$ we get equation (39) of \cite{SW}. Now the expansion about
${\bi x}={\bi x}'$ gives
\begin{eqnarray}
\fl G^L_t({\bi x},{\bi x}')=-\frac{1}{|{\bi x}-{\bi x}'|}\frac{1-m/2\varrho '}{(1+m/2\varrho ')^3}\nonumber\\
\lo\times\left[ 1+\frac{-\varrho
'+m(1-m/4\varrho ')}{\varrho '^3(1+m/2\varrho ')(1-m/2\varrho ')}\,{\bi x}'({\bi x}-{\bi x}')+\Or ({\bi x}-{\bi
x}')^2\right] \nonumber\\
\lo-\frac{m}{\varrho '^2}\frac{1}{(1+m/2\varrho ')^4}\left[ 1-\frac{1}{\varrho '^2}\frac{1-m/2\varrho '}{1+m/2\varrho
'}\,{\bi x}'({\bi x}-{\bi x}')+\Or ({\bi x}-{\bi
x}')^2\right] \;. \label{exGtL}
\end{eqnarray}
In equation \eref{empken} is $\sqrt{-g({\bi x}')}\,\delta^0=\delta^3({\bi x'}-{\bi x}'')$. Like for $G^L_t({\bi x},{\bi
x}')$, we build a ball around the point ${\bi x}'$. Both, ${\bi x}''$ and ${\bi x}$, are in a neighbourhood of ${\bi
x}'$. So we can set ${\bi x}={\bi x}''$ in the expression of $\delta^0$ and substitute
\be \label{del}
\sqrt{-g({\bi x}')}\,\delta^0=\frac{1}{4\pi a^2}\lim_{a\rightarrow 0}\delta (|{\bi x}-{\bi x}'|-a)\;.
\ee
Using the expansion \eref{exGtL}, the integral \eref{empken} yields
\be \label{empken2}
U_{\rm em}^{\delta}=\frac{1}{2a}\frac{1-m/2\varrho '}{(1+m/2\varrho ')^3}+\frac{m}{2\varrho
'^2}\frac{1}{(1+m/2\varrho ')^4}\;.
\ee
Now, we have the result for $U_{\rm em}$ \eref{emen5}, which is valid for any charge distribution.

If we choose pointlike distributions such as point charges or a ring of charge (cf, section~\ref{tsex}), $j^0$ has the form
\be \label{ladvert}
\sqrt{-g}\,j^0=Q\,\delta^3({\bi x}-{\bi x}_0)\;,
\ee
where $Q$ is the total charge. Then, we obtain from the equations \eref{meco} and \eref{emen5}
\be \label{enint}
\fl U_{\rm mech}+U_{\rm
em}=-m_0\,\frac{1-m/2\varrho_0}{1+m/2\varrho_0}+\frac{Q^2}{2a}\frac{1-m/2\varrho_0}{(1+m/2\varrho_0)^3}+\frac{mQ^2}{2\varrho_0^2}\frac{1}{(1+m/2\varrho_0)^4}\;.
\ee
Now, in order to renormalize the mass $m_0$ to $M$, we must express the radius of the ball charge in local freely falling
coordinates. With $x^{\overline{\mu}}=\Lambda^{\overline{\mu}}_{\,\,\nu}\,x^\nu$ we obtain
\be \label{renoa}
\overline{a}=a(1+m/2\varrho_0)^2\;.
\ee
Then, we can write
\be \label{enint2}
U_{\rm mech}+U_{\rm em}=M\,\frac{1-m/2\varrho_0}{1+m/2\varrho_0}+\frac{mQ^2}{2\varrho_0^2}\frac{1}{(1+m/2\varrho_0)^4}\;,
\ee
with
\be \label{massrem}
M=-m_0+\lim_{\overline{a}\rightarrow 0}\frac{Q^2}{2\overline{a}}\;.
\ee
It follows from \eref{exfo2} that
\be \label{exkr}
F^{\bar{i}}_{\rm
ext}=\frac{Mm\,x_0^i}{\varrho_0^3}\frac{1}{(1+m/2\varrho_0)^3}\frac{1}{1-m/2\varrho_0}-\frac{m\,Q^2x_0^i}{\varrho_0^4}\frac{1}{(1+m/2\varrho_0)^6}\;.
\ee
For ${\bi x}_0=(0,0,b)$ equation \eref{exkr} is in agreement with equation (75) of \cite{SW}. In Schwarzschild
coordinates \eref{exkr} becomes
\be \label{exkr2}
F^{\bar{i}}_{\rm ext}=\left( \frac{Mm}{r_0^2}\left( 1-\frac{2m}{r_0}\right) ^{-1/2}-\frac{m\,Q^2}{r_0^3}\right)\left(
\begin{array}{c}
      \sin\vartheta_0\cos\varphi_0 \\
      \sin\vartheta_0\sin\varphi_0 \\
      \cos\vartheta_0
\end{array} \right) \;.
\ee
\Eref{exkr} (and \eref{exkr2}) holds only for pointlike distributions and if we assume for $U_{\rm mech}$ an ideal fluid
model. However, the repulsive self-force depends not on the chosen mechanical model, but only on the chosen charge
distribution. Thus, the electrostatic self-force for pointlike test distributions, i.e. the second term on the right-hand side of equation \eref{exkr} or
\eref{exkr2}, is always (in the freely falling system)
\be \label{exkr3}
F^{\overline{r}}_{\rm ext}=-\frac{m\,Q^2}{r_0^3}\;,
\ee
where $F^{\overline{r}}_{\rm ext}$ is the radial component of the self-force in terms of the pointlike test
distribution's Schwarzschild radial coordinate. The force is radially directed because of the spherical symmetry of the
spacetime and depends only on the total charge $Q$.

\appendix
\section*{Appendix}
\setcounter{section}{1}
We generalize the wellknown Green's identities for $p$-forms on a (pseudo-) Riemannian manifold. Loomis and Sternberg
\cite{LS} and Flanders \cite{HF} give the derivation for functions on the $n$-dimensional Euclidean space. The result
for Riemannian manifold can be found in section~21 of Holmann and Rummler's book \cite{HR2}. Thirring also presents a
version of Green's second identity \cite{WT}. However, his result is not well adapted to our case. Since the Schwarzschild spacetime is pseudo-Riemannian, we have to generalize the formula in \cite{HR2}.

Let $(M,g)$ be an $n$-dimensional oriented (pseudo-) Riemannian manifold and let $D$ be a region of $M$ with smooth boundary such that $\bar{D}$ is compact. For the two forms $\alpha\in\bigwedge_{p}(M)$ and $\beta\in\bigwedge_{q}(M)$ the anti-Leibniz rule gives
\be \label{aLr}
\rmd (\alpha\wedge\beta )=\rmd\alpha\wedge\beta +(-1)^{p}\,\alpha\wedge \rmd\beta\;.
\ee
Now, let $u,v\in\bigwedge_{p}(M)$ be two $p$-forms. Then, with (\ref{aLr}), we obtain
\be \label{aLruv}
\rmd (u\wedge\ast\,\rmd v)=\rmd u\wedge\ast\,\rmd v+(-1)^{p}\,u\wedge \rmd\ast \rmd v\;.
\ee
Using Stokes theorem we obtain
\be \label{aLrS}
\fl\int_{\partial D}u\wedge\ast\,\rmd v=\int_{D}\rmd (u\wedge\ast\,\rmd v)=\int_{D}\rmd u\wedge\ast\,\rmd v+(-1)^{p}
\int_{D}u\wedge \rmd\ast \rmd v\;.
\ee
Note that $\rmd u\wedge\ast\,\rmd v=\rmd v\wedge\ast\,\rmd u$. If we write down (\ref{aLrS}) again with $u$ and $v$ interchanged, and then subtract it from (\ref{aLrS}), we have
\be \label{sGia}
\int_{\partial D}(u\wedge\ast\,\rmd v-v\wedge\ast\,\rmd u)=(-1)^{p}
\int_{D}(u\wedge \rmd\ast \rmd v-v\wedge \rmd\ast \rmd u)\;.
\ee

Now, we write the right-hand side of (\ref{sGia}) in a different form. For every form $\omega\in\bigwedge_{k}(M)$ (see \cite{NS})
\be \label{nsa}
\ast\ast\omega =(-1)^{k(n-k)}\,{\rm sgn}(g)\,\omega\;.
\ee
This gives us
\be \label{nsb}
(-1)^{k(k-n)}\,{\rm sgn}(g)\,\ast\ast\,\omega =\omega\;.
\ee
With $\omega =\rmd\ast \rmd v$ and $k=n-p$
\begin{eqnarray} \label{nsc}
\rmd\ast \rmd v & = & (-1)^{(n-p)(n-p-n)}\,{\rm sgn}(g)\,\ast\ast\,\rmd\ast \rmd v \nonumber\\
 & = & (-1)^{p(p-n)}\,{\rm sgn}(g)\,\ast\ast\,\rmd\ast \rmd v\;.
\end{eqnarray}
The codifferential $\delta :\bigwedge_{q}(M)\rightarrow\bigwedge_{q-1}(M)$ is defined by \cite{NS}
\be \label{nsd}
\delta := {\rm sgn}(g)\,(-1)^{nq+n}\,\ast \rmd\ast\;.
\ee
We solve (\ref{nsd}) for $\ast \rmd\ast$
\be \label{nse}
\ast \rmd\ast =(-1)^{-nq-n}\,{\rm sgn}(g)\,\delta\;.
\ee
Setting (\ref{nse}) in (\ref{nsc}) we obtain for $\rmd\ast \rmd v$ (and in analogous way for $\rmd\ast \rmd u$) $(q=p+1)$
\be \label{nsf}
\rmd\ast \rmd v = (-1)^{p}\,\ast\delta \rmd v\,,\quad \rmd\ast \rmd u = (-1)^{p}\,\ast\delta \rmd u\;.
\ee
We set (\ref{nsf}) in (\ref{sGia})
\be \label{sGib}
\int_{\partial D}(u\wedge\ast\,\rmd v-v\wedge\ast\,\rmd u)=\int_{D}(u\wedge\ast\,\delta \rmd v-v\wedge\ast\,\delta \rmd u)\;.
\ee
Since for two $p$-forms $\alpha ,\beta$
\[
\alpha\wedge\ast\,\beta =\beta\wedge\ast\,\alpha\;,
\]
equation (\ref{sGib}) becomes
\be \label{sGic}
\int_{\partial D}(u\wedge\ast\,\rmd v-v\wedge\ast\,\rmd u)=\int_{D}(\delta \rmd v\wedge\ast\,u-\delta \rmd u\wedge\ast\,v)\;.
\ee

Now, we consider the combination
\[
\delta u\wedge\ast\,v\in\bigwedge{}_{n-1}(M)\;.
\]
Using the anti-Leibniz rule and Stokes theorem again we have
\be \label{aLrSa}
\int_{\partial D}\delta u\wedge\ast\,v=\int_{D}\rmd\delta u\wedge\ast\,v+(-1)^{p-1}\int_{D}\delta u\wedge \rmd\ast\,v\;.
\ee
We interchange $u$ and $v$ and subtract the new equation from (\ref{aLrSa})
\begin{eqnarray}
\fl \int_{\partial D}(\delta u\wedge\ast\,v-\delta v\wedge\ast\,u) & = & \int_{D}(\rmd\delta u\wedge\ast\,v-\rmd\delta v\wedge\ast\,u) \nonumber\\
 & & {}+(-1)^{p-1}\int_{D}(\delta u\wedge \rmd\ast v-\delta v\wedge \rmd\ast u)\;. \label{sGid}
\end{eqnarray}
Now (by definition (\ref{nsd}))
\begin{eqnarray}
\delta u\wedge \rmd\ast v & =& {\rm sgn}(g)\,(-1)^{np+n}\,\ast \rmd\ast u\wedge \rmd\ast v \nonumber \\
 & = & {\rm sgn}(g)\,(-1)^{np+n}\,\ast \rmd\ast v\wedge \rmd\ast u=\delta v\wedge \rmd\ast u\;. \label{nsg}
\end{eqnarray}
If we set (\ref{nsg}) into (\ref{sGid}), the two last terms on the right-hand side of (\ref{sGid}) cancel and we find
\be \label{sGie}
\int_{\partial D}(\delta v\wedge\ast\,u-\delta u\wedge\ast\,v)=\int_{D}(\rmd\delta v\wedge\ast\,u-\rmd\delta u\wedge\ast\,v)\;.
\ee
Now, we add (\ref{sGic}) and (\ref{sGie}) and obtain Green's second identity written with $p$-forms on a (pseudo-) Riemannian manifold
\be \label{sGi}
\fl\int_{\partial D}(\delta v\wedge\ast\,u-\delta u\wedge\ast\,v+u\wedge\ast\, \rmd v-v\wedge\ast\,\rmd u)=\int_{D}(\Box v\wedge\ast\,u-\Box u\wedge\ast\,v)\;,
\ee
where $\Box := \rmd\circ\delta +\delta\circ \rmd $ is the Laplace--Beltrami operator.


\Bibliography{50}

\bibitem{WI}Israel W 1968 {\it Commun. Math. Phys.} {\bf 8} 245--60
\item[]Israel W 1967 {\it \PR}{\bf 164} 1776--79

\bibitem{HR1}Hanni R S and Ruffini R 1973 {\it Black Holes}\, ed 
C DeWitt and B S DeWitt (New York: Gordon and Breach) p R 57--73
\item[]Hanni R S and Ruffini R 1973 {\it \PR}D {\bf 8} 3259--65

\bibitem{CW}Cohen J M and Wald R M 1971 {\it \JMP} {\bf 12} 1845--49

\bibitem{BL}Linet B 1976 {\it \JPA} {\bf 9} 1081--87

\bibitem{BC}Carter B 1973 {\it Black Holes} ed
C DeWitt and B S DeWitt (New York: Gordon and Breach) p 57

\bibitem{SH}Hawking S W 1972 {\it Commun. Math. Phys.} {\bf 25} 152--66

\bibitem{DR}Robinson D C 1974 {\it \PR}D {\bf 10} 458--60

\bibitem{GO}Ginzburg V L and Ozernoi L M 1964 {\it Zh. Eksp. Teor. Fiz.} {\bf 47} 1030--40 (Engl. transl. 1965 {\it
Sov. Phys.--JETP} {\bf 20} 689--96)

\bibitem{AC}Anderson J L and Cohen J M 1970 {\it Astrophys. Space Sci.} {\bf 9} 146--52

\bibitem{JP}Petterson J A 1974 {\it \PR}D {\bf 10} 3166--70

\bibitem{RP}Price R H 1972 {\it \PR}D {\bf 5} 2439--54

\bibitem{Cetal}de la Cruz V, Chase J E and Israel W 1970 {\it \PRL}{\bf 24} 423--26

\bibitem{RW}Wald R M 1972 {\it \PR}D {\bf 6} 1476--79

\bibitem{PG}Ghosh P 2000 {\it Mon. Not. R. Astron. Soc.} {\bf 315} 89--97
\item[](Ghosh P 1999 The structure of black hole magnetosphere: I. Schwarzschild black holes {\it Preprint}
astro-ph/9907427)

\bibitem{NS}Straumann N 1984 {\it General Relativity and Relativistic Astrophysics} (Berlin: Springer)

\bibitem{MO}Magnus W, Oberhettinger F and Soni R P 1966 {\it Formulas and Theorems for the Special Functions of
Mathematical Physics} (Berlin: Springer)

\bibitem{WW}Walter W 1990 {\it Gew\"ohnliche Differentialgleichungen} (Berlin: Springer) p 133

\bibitem{EC}Copson E T 1928 {\it \PRS}A {\bf 118} 184--94

\bibitem{DeW}DeWitt C M and DeWitt B S 1964 {\it Physics, NY} {\bf 1} 3

\bibitem{BG}Berends F A and Gastmans R 1976 {\it Ann. Phys., NY} {\bf 98} 225

\bibitem{CMac}MacGruder C H III 1978 {\it Nature} {\bf 272} 806

\bibitem{AV}Vilenkin A 1979 {\it \PR}D {\bf 20} 373

\bibitem{SW}Smith A G and Will C M 1980 {\it \PR}D {\bf 22} 1276--84

\bibitem{ZF1}Zel'nikov A I and Frolov V P 1982 {\it Zh. Eksp. Teor. Fiz.} {\bf 82} 321 (Engl. trans. 1982 {\it Sov. Phys.--JETP} {\bf
55} 191--98)

\bibitem{ZF2}Frolov V P and Zel'nikov A I 1980 {\it 9th Int. Conf. on General Relativity and Gravitation
(Jena)} vol~3, p~555

\bibitem{BC2}Carter B 1979 {\it General Relativity: An Einstein Centenary Survey} ed S W Hawking and W Israel (New York:
Cambridge University Press) p~359

\bibitem{LS}Loomis L H and Sternberg S 1990 {\it Advanced Calculus} (London: Jones and Bartlett) p 476

\bibitem{HF}Flanders H 1963 {\it Differential Forms with Applications to the Physical Sciences} (New York: Academic) p 83

\bibitem{HR2}Holmann H and Rummler H 1972 {\it Alternierende Differentialformen} (Mannheim: BI-Wissenschaftsverlag) p 232

\bibitem{WT}Thirring W 1997 {\it Classical Mathematical Physics} (New York: Springer) p 306
\endbib
\end{document}